\documentclass[12pt]{article}
\usepackage{epsf}
\usepackage{graphicx}
\usepackage{lscape}
\usepackage{amssymb}
\usepackage{pazh}
\tightenlines

\def\phflux{phot.$\cdot$cm$^{-2}$s$^{-1}$keV$^{-1}$}
\def\a{$^{\mbox{\small a}}$}
\def\b{$^{\mbox{\small b}}$}
\def\c{$^{\mbox{\small c}}$}
\def\d{$^{\mbox{\small d}}$}
\def\GX{\mbox{GX301-2}}

\def\etal{{\it et~al.}}

\def\fdg{\hbox{$.\!\!^\circ$}}
\begin{document}
 
{\footnotesize Astronomy Letters, Vol. 30, No. 8, 2004, pp. 540­548. 
Translated from Pis'ma v Astronomicheskii Zhurnal, Vol. 30, No. 8, 
2004, pp. 596­604.
Original Russian Text Copyright \copyright\, 2004 by Tsygankov, 
Lutovinov, Grebenev, Gilfanov, Sunyaev.}

%\hline
%\hline

\title{\bf Observations of the X-ray Pulsar GX 301-2 with the ART-P 
Telescope of the Granat Observatory}   
 
\author{\bf \hspace{-1.3cm}\copyright\, 2004 \ \
   S.S.Tsygankov\affilmark{1}, A.A.Lutovinov\affilmark{1}$^{\,*}$, 
  S.A.Grebenev\affilmark{1}, M.R.Gilfanov\affilmark{1,2},
   R.A.Sunyaev\affilmark{1,2}}     
 
\affil{
$^1$ {\it Space Research Institute, Russian Academy of Sciences, 
Profsoyuznaya ul. 84/32, Moscow 117810, Russia}\\
$^2$ {\it Max-Planck-Institut f\"ur Astrophysik, Karl-Schwarzschild-Str. 1, 
Postfach 1317, D-85741 Garching, Germany }}
 
\vspace{2mm}
%\received{~~~~~~~~}
 
\sloppypar 
\vspace{2mm}
\noindent

%\abstract

{ The variability of the X-ray flux from the pulsar GX 301-2 
is analyzed by using data from the
ART-P telescope of the Granat Observatory. The intensity variations 
with time scales of several thousand
seconds are studied at various orbital phases. The high-state 
flux from the source exceeds its low-state flux
by as much as a factor of 10. The hardness and spectrum of 
the source are shown to change greatly with
its intensity. These intensity variations are most likely 
caused by substantial inhomogeneities in the stellar
wind from the companion star. 

\copyright\, 2004 MAIK "Nauka/Interperiodica".}


\vfill
 
{$^{*}$ E-mail: aal@hea.iki.rssi.ru}
\newpage
\thispagestyle{empty}
\setcounter{page}{1}

\section*{INTRODUCTION}

    The X-ray pulsar GX 301-2, which forms a binary
system with the blue supergiant Wray 977 with an
eccentricity of  $\sim$0.462 and an orbital period of 
$\sim$41.5 (Sato et al. 1986), is one of the longest-period
pulsars known to date. Throughout the history of
its observations, the spin period of this source has
repeatedly underwent abrupt changes. In April 1984,
the Tenma satellite measured the maximum pulsation
period, 701.14 s (Sato et al. 1986), which decreased
sharply to $\sim$677 s in 1992 (Lutovinov et al. 1994;
Chichkov et al. 1995). At this point, however, the
period of prolonged steady spin-up with a mean rate
of \mbox{$\dot P/P \sim$ -4.4 x 10$^{-3}$~ yr$^{-1}$}~ ended, and the 
pulsation period stabilized at  $\sim$679.5 s in November 1996
(Pravdo et al. 2001). On smaller time scales, the
switches between spin-up and spin-down of the 
neutron star manifest themselves much more frequently
(Chichkov et al. 1995; Koh et al. 1997). The X-ray
flux recorded from the binary is variable during the
orbital cycle. As the pulsar approaches the optical
star, the X-ray luminosity of the binary increases and
reaches its maximum 1.2 days before periastron 
passage (White et al. 1984). An increase in the intensity
of the source is also observed near the apoastron
(Chichkov et al. 1995; Koh et al. 1997).
    
Parkes et al. (1980) determined the parameters
of the optical star (its radius  $\sim43R_{\odot}$ and mass 
$\sim$30M$_{\odot}$) and the distance to the binary system ($d\sim$ 
1.8 kpc). Note that the precise spectral type of the
companion star has not yet been firmly established.
Some of the authors classify it as a B star (Vidal 1973;
Kaper et al. 1995), while others (Parkes et al. 1980)
classify it as an emission-line Be star. The normal star
may occupy an intermediate position and occasionally
switches over from one state to another.

     At present, there is no agreement about the
structure and distribution of the matter in the 
binary GX 301-2/Wray 977 either. Based on the very
small changes in the characteristic values of these
parameters with pulse phase, Tashiro et al. (1991)
concluded that the distribution of the photoabsorbing
and iron line emitting matter was spherically 
symmetric. On the other hand, having performed detailed
phase-resolved spectroscopy, Leahy et al. (1990)
concluded that the distribution of the matter around
the neutron star was nonuniform. All authors point
out a significant hydrogen column density reaching
$\sim2$x$10^{24}$~ cm$^{-2}$.

     In this paper, we present the results of a 
comprehensive analysis of the observational data for the
pulsar GX 301-2 obtained with the ART-P telescope
of the Granat orbital observatory, identifying two 
distinct (high and low) states in its light curve.

\section*{OBSERVATIONS}

     The X-ray pulsar GX 301-2 was observed by
the Granat observatory from January 1991 through
February 1992. Over this period, the ART-P 
telescope conducted four observing sessions for this
source with a total useful time of $\sim97$ ks (see
Table 1).

    The ART-P X-ray telescope consists of four 
coaxial, completely independent modules, each of which
includes a position-sensitive detector with a 
geometrical area of $625$~cm$^{2}$ and a coded mask. The
telescope is sensitive to photons in the energy range
2.5-60 keV (the energy resolution is $\sim22$\% in the
$5.9$-keV calibration iron line) and can image the sky
within a 3\fdg4$\times$3\fdg6 field of view with a nominal angular
resolution of  $\sim5$ arcmin (the angular size of the mask
element). For a more detailed technical description of
the telescope, see Sunyaev et al. (1990).

    The observations were carried out in 
photon-by-photon mode in which the coordinates of each 
photon on the detector, its energy (1024 channels), and
arrival time (the photon arrival time was accurate
to within 3.9 ms, and the dead time was 580 ~$\mu$s)
were written into the ART-P buffer memory. This
mode allows both timing and spectral analyses of the
emission from each X-ray source within the ART-P
field of view to be performed. Data were transferred
to the main memory after the temporary buffer was
filled (once in 200-250 s--the exposure time) during
$\sim$25-30 s,which led to gaps in information.

    All of the observations of the pulsar GX 301-2
were performed by the third ART-P module with a
reduced sensitivity to soft (3-8 keV) energies. 
Subsequently, this complicated and, in several cases, did
not allow a spectral analysis of the X-ray emission.

\section*{TIMING ANALYSIS}

    Our studies showed that the flux from GX 301-2
near the periastron and apoastron is highly variable.
Two states, high and low, can be arbitrarily identified;
the switches between them can occur several times
during one observing session. Figure 1a shows the
light curves of the pulsar for all four observing 
sessions. The hardness, which is defined as the ratio of
the fluxes from the source in the energy ranges 30-40
and 10-20 keV, is given for each light curve (Fig. 1b).

    We see from the figure that the switches of the
source from the low state to the high state and back
are rapid; its intensity during such switches changes
by about an order of magnitude. At the same time,
the hardness of the source during the switch to its
high state decreases (by a factor of $\sim$ 3). Note that
although the absolute stability of the low-state 
intensity is higher than that of the high-state intensity,
as illustrated by Fig. 1, the fractional rms of these
variations are close. Thus, the fractional rms during
periastron passage changed from  $\sim$21\% to  $\sim$34\% in
the low and high states, respectively; near the 
apoastron, its values were $\sim$ 25\% and  $\sim$35\% for the low
and high states, respectively. Near an orbital phase of
 $\sim$0.3, the fractional rms was $\sim$34\%.

    Table 1 gives the pulsation periods determined by
a superposed-epoch analysis after the correction of
the photon arrival time for the motion of the neutron
star in the binary system and for the motion of the
Earth and the spacecraft. Throughout our 
observations (about 13 months), the period decreased by 
$\sim$6 s, which corresponds to a mean value of 
$\dot P/P \approx -8.2 \times 10^{-3}$~yr$^{-1}$.

\pagebreak

%\begin{landscape}
\begin{table}[h]
\centering
{\bf Table 1.} ART-P observations of the pulsar GX 301-2 during 1991-1992\a

\vspace{2mm}

\begin{tabular}{c|c|c|c|c|c} \hline \hline
Date & Orbital phase & Exposure time, s& Flux, mCrab& $L_X$,\b$10^{36}$ 
\mbox{\rm erg s$^{-1}$} & Period, s \\ \hline
&&&&\\ [-4mm]
09.01.91 & 0.03-0.04 & 23584 & & &$682.83\pm0.04$\\[1mm]
(I)   & & 2835 & $~32\pm8~$  & $0.29\pm0.07$&\\[1mm]
(II)  & & 6126 & $301\pm8~$  & $2.72\pm0.08$&\\[1mm]
(III) & & 7067 & $126\pm6~$  & $1.14\pm0.05$&\\[1mm]
(IV)  & & 7556 & $420\pm9~$  & $3.81\pm0.08$&\\[2mm]
13.08.91 & 0.23-0.24 & 33346 & --\c & --\c &$678.91\pm0.43$\\[2mm]
29.01.92 & 0.31 & 11376 & $~28\pm5$ & $0.26\pm0.05$&$678.77\pm1.82$ \\[2mm]
07.02.92 & 0.52-0.53 & 28540 &&&$676.83\pm0.08$\\[1mm]
(I)   & & 986   & $198\pm21$  & $1.80\pm0.19$&\\[1mm]
(II)  & & 13965 & $~37\pm5~$  & $0.33\pm0.04$&\\[1mm]
(III) & & 13589 & $167\pm6~$  & $1.51\pm0.06$&\\[1mm]
\hline
\end{tabular}

\vspace{3mm}

\begin{tabular}{cl}

\a& In the energy range 6-40 keV.\\
\b& For the assumed distance to the source of $d=1.8$ kpc.\\
\c& For technical reasons, the spectrum of the source cannot be reconstructed.\\

\end{tabular}
\end{table}
%\end{landscape}

    We analyzed the behavior of the pulse profile and
the pulse fraction for the source as a function of
its state and the energy range. Figure 2 shows the
phase light curves of the pulsar averaged over each
of the sessions in five energy ranges (3-6, 6-10, 10-
20, 20-30, and 30-40 keV). The pulse profile has a
double-peaked shape with a slight dominance of the
first peak; the intensity of the second peak increases
relative to the first peak as the energy increases. 
During the source's switches from one state to the other,
the pulse shape remains virtually unchanged, only the
intensity of the emission changes.
 
  The pulse fraction, which is defined as 
$P=(I_{max}-I_{min})/(I_{max}+I_{min})$, where 
$I_{max}$ and $I_{min}$ are
the background-corrected count rates at the 
maximum and minimum of the pulse profile, is given in
Table 2 for all observing sessions in various energy
ranges (6-10, 10-20, 20-30, 30-40, and 6-40 keV)
as a function of the source's state. In view of the
technical peculiarities of the third module described
above, the pulse fraction in the energy range 3-6 keV
was not calculated.

   It follows from the table that the pulse fraction
remains fairly high and weakly depends on the energy
range and the source's intensity, although the 
pulsations in the energy range 30-40 keV are slightly
blurred and the pulse fraction slightly decreases in
the energy range 10-20 keV. As the source passed
through the apoastron (February 7, 1992), its 
pulsations completely disappeared when it was in the low
state. For this case, the table gives only upper limits
on the pulse fraction.

\begin{table}[h]
\noindent
\centering
{\bf Table 2. }{ Pulse fractions for \GX.}\\   
\centering
\vspace{1mm}
\small{

\begin{tabular}{c|c|c|c|c|c}
\hline\hline
Date &\multicolumn{5}{c}{Pulse fraction, \%}\\  \cline{2-6}
     &6-10 keV&10-20 keV&20-30 keV&30-40 keV&6-40 keV \\
\hline
09.01.91&&&&&\\
I&---\a&---\a&---\a&---\a&---\a\\
II&$60.9\pm4.9$&$55.2\pm2.1$&$61.5\pm3.4$&$35.9\pm7.1$&$54.9\pm1.8$\\
III&$78.5\pm10.6$&$63.0\pm3.3$&$83.6\pm8.9$&$35.6\pm7.2$&$64.2\pm3.2$\\
IV&$36.6\pm2.2$&$32.6\pm1.3$&$43.3\pm2.9$&$33.9\pm6.6$&$34.2\pm1.1$\\
13.08.91&$58.6\pm7.7$&$47.4\pm3.2$&$50.6\pm7.9$&$20.9\pm6.3$&$44.7\pm3.0$\\
29.01.02&$91.1^{+ 8.9}_{-38.9}$&$76.8\pm13.4$&---\b&---\b&$82.8\pm15.4$\\
07.02.92&&&&&\\
I&---\a&---\a&---\a&---\a&---\a\\
II&$44.3$\c&$19.9$\c&$24.9$\c&$15.5$\c&$15.2$\c\\
III&$44.7\pm3.4$&$53.5\pm1.9$&$61.4\pm4.1$&$82.1\pm24.8$&$53.3\pm1.8$\\

\hline
\end{tabular}

\vspace{3mm}

\begin{tabular}{cl}

\a& Cannot be determined because of poor statistics.\\
\b& Cannot be determined for technical reasons.\\
\c& The $3\sigma$ upper limit.
\end{tabular}}
\vspace{3mm}
\end{table}

\pagebreak

\section*{SPECTRAL ANALYSIS}

    To study the properties of the pulsar GX 301-2 in
detail, we performed a spectral analysis of its emission
in different states. The main model used in our 
analysis was a simple power law, which is most 
characteristic of X-ray pulsars at energies up to $\sim$ 10-15 keV.
However, we did not always use this simple model. As
required, we used various modifications of the model
that were specified in general form by the equation

\begin{equation}
I(E)=I_{10}\,\biggl( \frac {E}{10\ \mbox{keV}} \biggr)^{-\alpha} exp(-\sigma_{A} N_{\rm H})\times  
\left\{\begin{array}{c} 1, 
\mbox{ if $E<E_{c}$};\\ \exp{[-(E-E_{c})/E_{f}\,]}, 
\mbox{if}\ E\geq E_{c},\\
\end{array}\right.
\end{equation}

where $E$~ is the photon energy in keV, $I_{10}$~ is the 
normalization of the power-law component to 10 keV, 
~$\alpha$~ is the photon spectral index, $E_{c}$~ is the cutoff energy,
$E_{f}$~ is the e-folding energy in the source's spectrum,
$N_{H}$~ is the hydrogen column density, and $\sigma_{A}(E)$~ is the
interstellar absorption cross section.

    This model includes both the low-energy cutoff
attributable to photoabsorption and the purely 
empirical multiplicative component that is commonly
used to describe the spectra of X-ray pulsars, the
high-energy cutoff (White et al. 1983). Formally,
photoabsorption was recorded only in one session on
January 9, 1991, when the source was in the low
state. In this case, the hydrogen column density was
$N_{H}=(7.9\pm3.2)\times10^{23}$ cm$^{-2}$. Such a high value
agrees well with the results by Haberl (1991), who
performed a detailed analysis of the photoabsorption
at various orbital phases. We failed to perform a more
detailed analysis of this parameter due to the technical
problems of the third ART-P module discussed above
and obtained only upper limits for the remaining 
observing sessions (see Table 3).

  Based on the $\Delta \chi^{2}$ test, we passed to a more 
complex model. The model was assumed to be acceptable
if the probability that the $\chi^{2}$ value did not improve
by chance exceeded 95\%. Table 3 gives the best-fit
parameters for the pulsar's spectra averaged over the
period of the observing session under consideration
that corresponded a certain state of the source.

    As we see from Table 3, the high-state spectrum
of the source is slightly softer than its low-state 
spectrum. The cutoff energy $E_{c}$ as well as the e-folding
energy $E_{f}$ remain constant, within the error limits,
and are in good agreement with the results of other
authors (Leahy et al. 1990). However, they show up
only in the high state, which may be attributable to
poor statistics when analyzing the emission from the
source in its low state. Figure 3 shows how the energy
spectrum changes with the source's state.

\begin{table}[t]
\centering

\hspace{-3mm}{\bf Table 3. }{Best-fit parameters for the spectrum of 
 GX301-2\a}

\vspace{2mm}

\hspace{-5mm}\begin{tabular}{@{}c|c|c|c|c|c|c}
\hline\hline
&&&&&\\ [-4mm]
 Date & $I_{10}$\b$,\times10^{-3}$ & $\alpha$ & $E_{c}$ & $E_{f}$
 &$N_{\rm H}, 10^{23}$ cm$^{-2}$&$\chi^{2}_{\,N} (N)$\c \\
\hline
 &&&&&\\ [-4mm]

09.01.91 &&&& \\
(I)   &  $1.26\pm1.06$ & $0.32\pm0.73$ &  &  &$8.4$\d &1.11(7)\\
(II)  & $15.76\pm0.75$ & $0.85\pm0.10$ & $22\pm3$ & $23\pm13$ &$0.4$\d& 1.19(17)\\
(III) & $13.15\pm3.40$ & $1.77\pm0.34$ &  &  &$7.9\pm3.2$& 1.73 (15)\\
(IV)  & $21.02\pm0.76$ & $0.71\pm0.06$ & $28\pm2$ & $11\pm5~$ &$0.3$\d& 2.72(13)\\[2mm]
29.01.92 & $1.66\pm0.97$ & $1.53\pm0.65$ &  &  &$2.9$\d& 0.21(5)\\[2mm]
07.02.92 &&&&& \\
(I)   &  $9.10\pm2.13$ & $0.76\pm0.31$ &  &  &$3.0$\d& 0.90(8)\\
(II)  &  $1.46\pm0.69$ & $0.59\pm0.39$ &  &  &$4.7$\d& 1.18(8)\\
(III) & $10.22\pm0.77$ & $1.04\pm0.24$ & $16\pm3$ & $31\pm13$ &$0.7$\d& 0.79(16)\\[1mm]
\hline
\end{tabular}

\vspace{3mm}

\begin{tabular}{cl}
\a& All errors are given at the  1$\sigma$ level.\\
\b& The flux at  10 keV (\phflux).\\
\c& The $\chi^2$ value normalized to the number of degrees of freedom  $N$.\\
\d& The $1\sigma$ upper limit on the atomic hydrogen column density.\\

\end{tabular}
\end{table}

\section*{DISCUSSION}

    Despite the abundance of observational data,
the precise accretion pattern in the binary system
GX 301-2/Wray 977 that would account for the 
nature of its X-ray emission is not yet completely clear.
The source exhibits variable emission, it behaves in
a new fashion from orbit to orbit, and the intensity
peaks and minima shift, occasionally replacing one
another (except for the peaks near the periastron
and apoastron, although their intensities also vary
significantly).

    Several models of the interaction of a compact
object with interstellar matter in a binary system were
suggested to interpret the observational data. Pravdo
et al. (1995) showed that the periodic outbursts were
difficult to explain only by isotropic accretion from the
stellar wind and explained this behavior of the source
either by the interaction of the neutron star with an
equatorially enhanced stellar wind typical for $Be$ stars
or by the formation of an accretion disk around the
neutron star, which may also be a source of angular
momentum, causing it to spin up. Leahy et al. (2002)
presented the results of a long-term monitoring of
GX 301-2 and considered several possible models
for the generation of emission, with the wind+stream
model yielding the best fit to the experimental data.

    The hydrodynamic simulations of a 
nonaxisymmetric gas stream in the binary system GX 301-
2/Wray 977 by Taam and Fryxell (1989) revealed
significant aperiodic luminosity fluctuations (with a
factor as large as 10) on time scales of the order
of several thousand seconds. Tashiro et al. (1991)
studied the aperiodic intensity variations of GX 301-2
that could be the result of plasma turbulence near the
neutron-star surface and that manifest themselves on
time scales up to several tens of seconds.

    The intensity variations considered here are also
aperiodic in pattern and are observed on time scales
from several tens of seconds to several hours (see
Fig. 1). The observed intensity fluctuations of the
source under study on short time scales are most
likely attributable to local inhomogeneities in the 
stellar wind, its clumpy structure. Of particular interest is
the observing session near the apoastron on 
February 7, 1992, when a prolonged decrease in the 
intensity of the source and the disappearance of pulsations
were recorded (part II of the last light curve in Fig. 1).

%\pagebreak
%\begin{landscape}

%\end{landscape}

%\pagebreak

A plausible explanation of the observed intensity
decrease is the eclipse of the neutron star by a 
substantial inhomogeneity in the stellar wind on the line
of sight between the observer and the object. Let
us attempt to estimate its characteristic parameters.
Following Castor et al. (1975), we can write

\begin{equation}
v_{w}(r)=v_{\infty}\biggl(1-\frac {R_{c}}{r} \biggr)^{\beta},  
\end{equation}

where $v_{\infty}$ is the terminal velocity of the stellar wind,
$R_{c}$ is the radius of the optical companion, and $\beta=0.5$.
Assuming the characteristic time the source
stays in its low state to be $\sim$ 10 ks and the range
of possible wind terminal velocities to be from $v_{\infty}=400$
km s$^{-1}$ (Kaper et al. 1995) to 1000 km s$^{-1}$
(Parkes et al. 1980), we obtain the size of the
inhomogeneity $l=(5\div12)R_{\odot}$ from formula (2). In our
case, the terminal velocity of the stellar wind was
calculated near the apoastron, which corresponds to
a distance of $\sim$ 5$R_{c}$ from the optical companion (Sato
et al. 1986). Since the orbital velocity of the neutron
star,  $\le$100 km s$^{-1}$, is low compared to the terminal
velocity of the wind, we disregarded it. Note that the
source switches to and from its low state not abruptly,
but over several (4-5) exposures (Fig. 1); i.e., the
duration of this switch is $\sim$ 800--1000 s. Thus, the
boundaries of the inhomogeneity are slightly blurred
and are $\sim$ 15--20\% of its total size.

     Figure 4 shows the pulse-height spectra of the
pulsar GX 301-2 obtained for its low and high states
during the observations on February 7, 1992. The
normalization of the low-state spectrum was 
multiplied by a factor of 2.6 to match it to the high-state
spectrum at energies above  $\sim$ 25 keV, where these
spectra have approximately the same shape. We see
that the low-state spectrum in softer energy channels
exhibits a cutoff. Taking into account this spectral
feature attributable to photoabsorption, we can 
estimate the atomic hydrogen column density at the time
in question, $N_{H}\sim1.7\times10^{24}$ atoms cm$^{-2}$. Assuming
the size of the inhomogeneity to be $l\sim10R_{\odot}$, we
determined its electron density, 
$n=\frac {N_{H}} {l}\sim2.4\times10^{12}$~cm$^{-3}$.

For comparison, Fig. 5 shows the pulse-height
profiles of the pulsar GX 301-2 at orbital phase 0.03--
0.04 near the periastron. We see that the shape of
the spectra is almost constant and does not depend
on the flux. Only their normalization changes, which
is indicative of the absence of enhanced (additional)
photoabsorption in the low state (although the upper
limits given in Table 3 are significant, which is most
likely due to the shortage of data in the soft spectral
range).

The dipping in the source's spectrum at high 
energies observed at the apoastron may be attributable
to Thomson scattering. The intensity ratio of 2.6 (see
above) implies that the scattering optical depth of the
condensation under consideration is  $\sim$1. The 
corresponding electron density of the inhomogeneity is
$n=\frac {\tau}{l\cdot\sigma_{T}}\sim2\times10^{12}$ ~cm$^{-3}$~ 
($\sigma_{T}$ is the Thomson scattering 
cross section) at a size of ~$l\sim10R_{\odot}$, which is in
agreement with the above densities of the cloud 
estimated from the observed absorption at low energies.
               
  Based on a model with an equatorially enhanced
stellar wind, Waters et al. (1988) showed for several
X-ray binary systems with $Be$ stars that its density
around such stars varies as

\begin{equation}
n(r)=n_{0}\biggl( \frac {r} {R_{c}} \biggr)^{-\gamma},  
\end{equation}

where $n_{0}$ is the gas density near the surface of the
optical star, $r$ is the distance from the star, and $\gamma$ is
a parameter equal to $\sim$ 3 for most of the stars under
consideration.   

          Using the values of $n_{0}$ from the same paper
($\sim10^{-11}$~g cm$^{-3}$) and formula (3), we obtain the
mean density of the stellar wind in the equatorial disk
around the optical companion near the apoastron,
$n(5R_{c})\sim8\times10^{-14}$~g cm$^{-3}$, which corresponds to
an electron density of $\sim5\times10^{10}$~ cm$^{-3}$.

Our estimates indicate that the inhomogeneities
of the scale under consideration are denser than the
average equatorial stellar wind characteristic of stars
of this spectral type by a factor of about 40-50.

    Note that the disappearance of pulsations is rather
difficult to explain in terms of our simple model in
which the neutron star is eclipsed by an 
inhomogeneity in the stellar wind with an optical depth $\tau\sim1$.
Following Mendoz and van der Klis (2000), we can
roughly estimate the optical depth of a cloud with
a size of ~$l\sim10R_{\odot}$ required to decrease the pulse
fraction by a factor of  $\sim$3 (see Table 2),  $\tau\sim4$. The
derived spread in optical depths may be attributable
to poor statistics when searching for pulsations and
determining the pulse fraction in the low state 
(particularly in soft energy channels).

  In conclusion, we present the following interesting
fact: the hardness of the source during periastron
and apoastron passage is proportional to its intensity
(Fig. 6) and can be formally described by the relation
$H\varpropto F^{-\beta}$, where $H$ is the hardness, and $F$ is the flux
from the source in mCrab. For the apoastron, this 
dependence is stronger ($\beta=-1.02\pm0.06$, filled circles)
than it is for the periastron ($\beta=-0.47\pm0.08$, open
circles).

\section*{ACKNOWLEDGMENTS}

   This study was supported by the Russian 
Foundation for Basic Research (projects no. 02-02-17347 and 04-02-17276)
and the Nonstationary Phenomena in Astronomy
Program. We wish to thank K.G. Sukhanov, flight
director, the staffs of the Lavochkin Research and
Production Center, RNIIKP, and the Deep Space
Communications Center in Evpatoria, the 
Evpatoria team of the Space Research Institute (Russian
Academy of Sciences), the team of I.D. Tserenin,
and B.S. Novikov, S.V. Blagii, A.N. Bogomolov,
V.I. Evgenov, N.G. Khavenson, and A.V. D'yachkov
from the Space Research Institute who operated the
Granat observatory, provided the scientific planning
of the mission, and performed a preliminary telemetry
data processing. We also wish to thank the team of
M.N. Pavlinsky (Space Research Institute) and the
staff of the former Research and Development Center
of the Space Research Institute in Bishkek who
designed and manufactured the ART-P telescope.

 \pagebreak
%****************************************************************

\section*{REFERENCES}
\parindent=0mm

1. Castor J.I., Abbott D.C., Klein R.I.,
 \apj\ {\bf 195}, 157 (1975).

2. Chichikiv M.A., Sunyaev R.A., Lapshov I.Yu. \etal, 
Astronomy Letters {\bf 21}, 435 (1995).

3. Haberl F., \apj\ {\bf 376}, 245 (1991).

4. Kaper L., Lamers H.J.G.L.M., Ruymaerkers 
E. \etal, \aap\ {\bf 300}, 446 (1995).

5. Koh D., Bildsten L., Chakrabarty D. \etal,
 \apj\ {\bf 479}, 993 (1997).

6. Leahy D.A., Matsuoka M., \asr\ {\bf 10}, (2)95 (1990).

7. Leahy D.A.,  \aap\ {\bf 391}, 219 (2002).

8. Lutovinov A.A., Grebenv S.A., Sunyaev R.A.\etal, 
Astronomy Letters {\bf 20}, 538 (1994).

9. Melatos A., Johnston S., Melrose D.B.,
 \mnras\ {\bf 275}, 381 (1995).

10. Mendoz M., van der Klis M., \mnras\ {\bf 318}, 938 (2000).

11. Parkes G.E., Mason K.O., Murdin P.G. \etal, 
 \mnras\ {\bf 191}, 547 (1980).

12. Pravdo S.H., Day C.S.R., Angelini L. \etal,
 \apj\ {\bf 454}, 872 (1995).

13. Pravdo S.H., Ghosh P., 
 \apj\ {\bf 554}, 383 (2001).

14. Sato N., Nagase F., Kawai N. \etal,
 \apj\ {\bf 304}, 241 (1986).

15. Sunyaev R.A., Babichenko S.I., Goganov D.A. \etal,
  \asr\ {\bf 10}, (2)233 (1990).

16. Taam R.E., Fryxell B.A., \apj\ {\bf 339}, 297 (1989).

17. Tashiro M., Makishima K., Ohashi T. \etal,
 \mnras\ {\bf 252}, 156 (1991).

18. Vidal N.V.,  \apj\ {\bf 186}, 81 (1973).

19. Waters L.B.F.M., Taylor A.R., van den
Heuvel E.P.J. \etal, \aap\ {\bf 198}, 200 (1988).

20. White N., Swank J., Holt S., \apj\ {\bf
  270}, 771 (1983).   

21. White N.E., Swank J.H., \apj\ {\bf 287}, 856 (1984).

\pagebreak
%************************************************************************
 
\newpage

\begin{landscape}
\begin{figure*}[t]
\includegraphics[angle=90,width=21cm,bb=40 20 550 790]{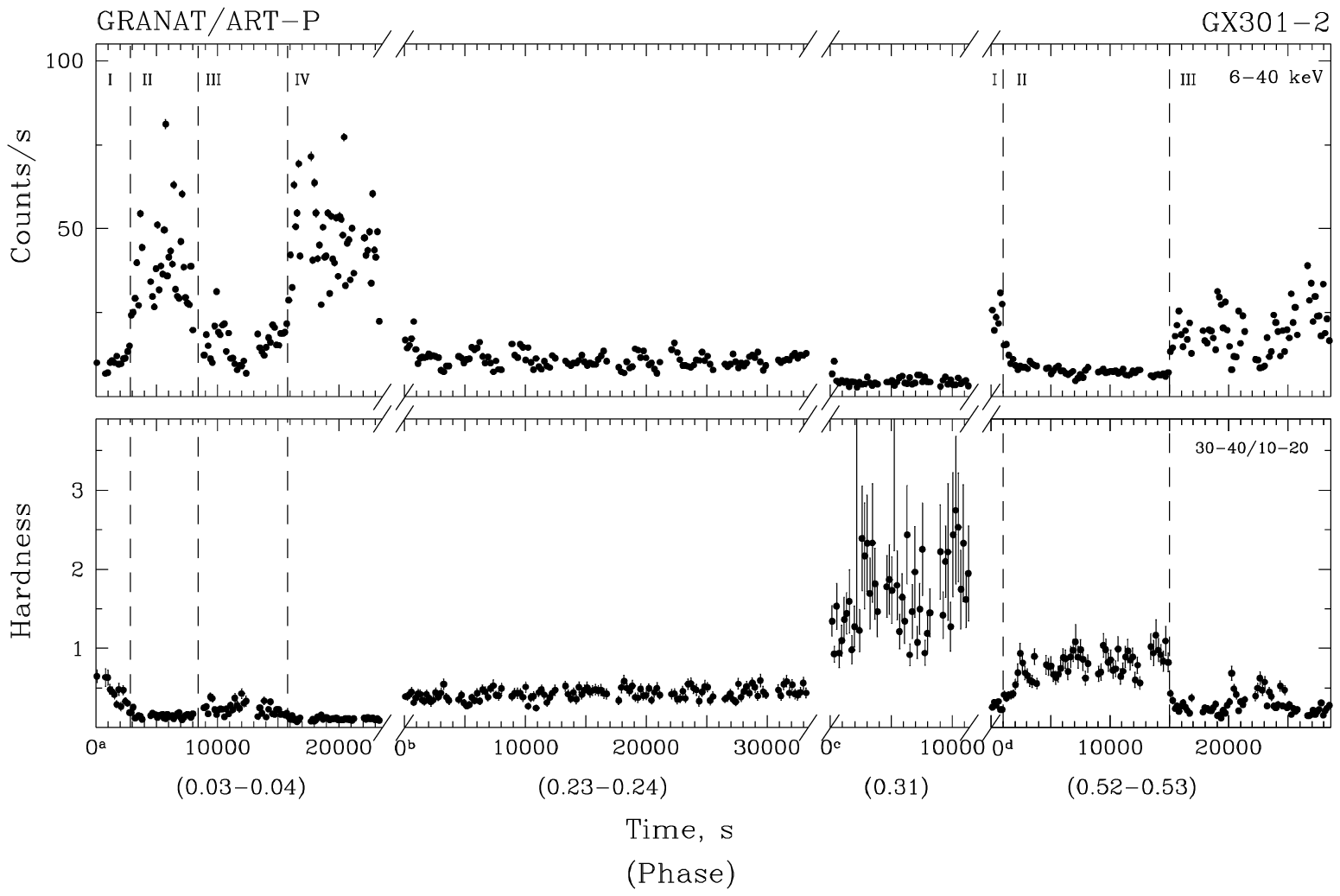}

\vfill

\caption{ 
 Light curve for the pulsar GX 301-2 (a) 
in the energy range 6-40 keV for four observing 
sessions and the corresponding
variations in the hardness of the source (b)--the ratio 
of the count rates in the energy ranges 30-40 and 10-20 keV
(the background was subtracted). The dashed lines mark the 
boundaries of the low and high states. Time in seconds
from the beginning of the observing session is along the 
horizontal axis. Zeros correspond to the following times: $^{a}$--UT 
18h34m05.502s (January 9, 1991); $^{b}$--UT 09h57m32.519s 
(August 13, 1991); $^{c}$--UT 17h15m16.208s (January 29, 1992); $^{d}$--
UT 18h16m06.201s (February 7, 1992). 
The corresponding orbital phase is given in parentheses 
for each session. The errors
correspond to one standard deviation.
}
\end{figure*}
\end{landscape}

\newpage

\begin{landscape}
\begin{figure*}[t]
\includegraphics[angle=90,width=25cm,bb=110 20 520 750,clip]{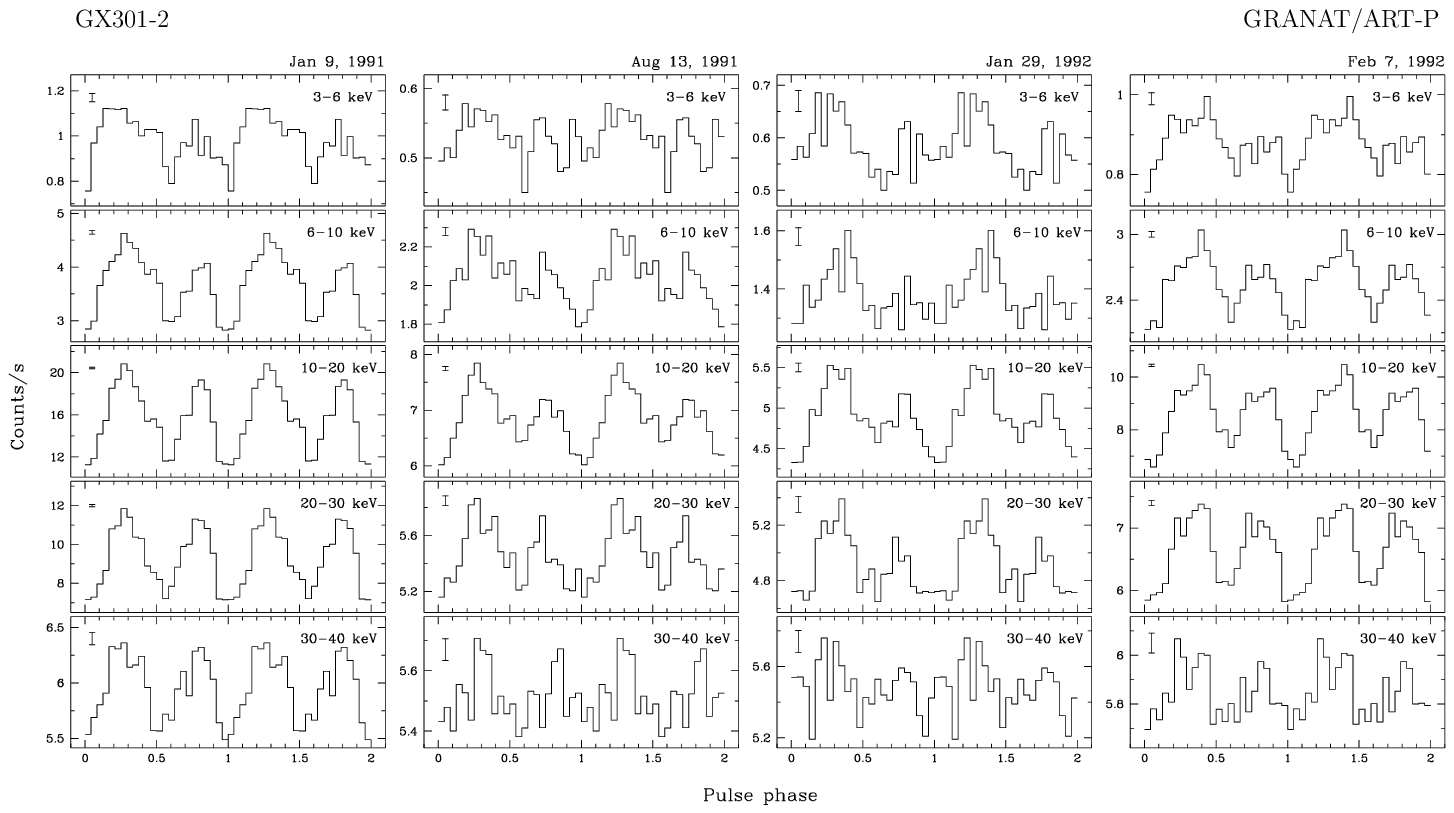}

\vfill

\caption{
ART-P pulse profiles for GX 301-2 in various energy ranges. 
The errors correspond to one standard deviation.}

\end{figure*}
\end{landscape}

\newpage

\begin{figure*}[t]
\includegraphics[width=14cm,bb=100 120 450 700,clip]{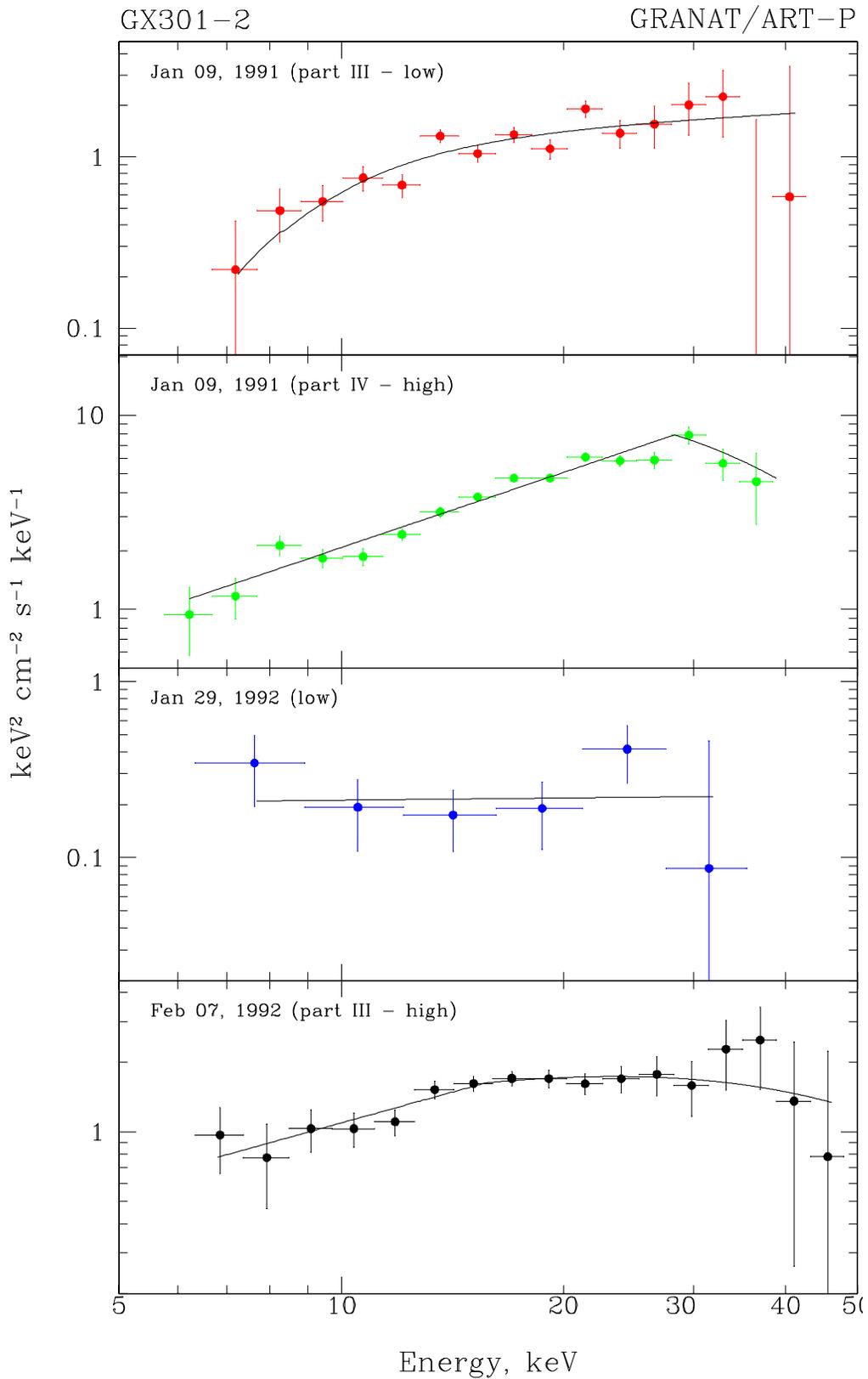}

\vfill

\caption{ART-P energy spectra for GX 301-2. The solid
lines represent the model fits to the spectra (see Table 3).
The errors correspond to one standard deviation.}

\end{figure*}

\newpage

\begin{figure*}[t]
\includegraphics[width=14cm,bb=90 140 550 700]{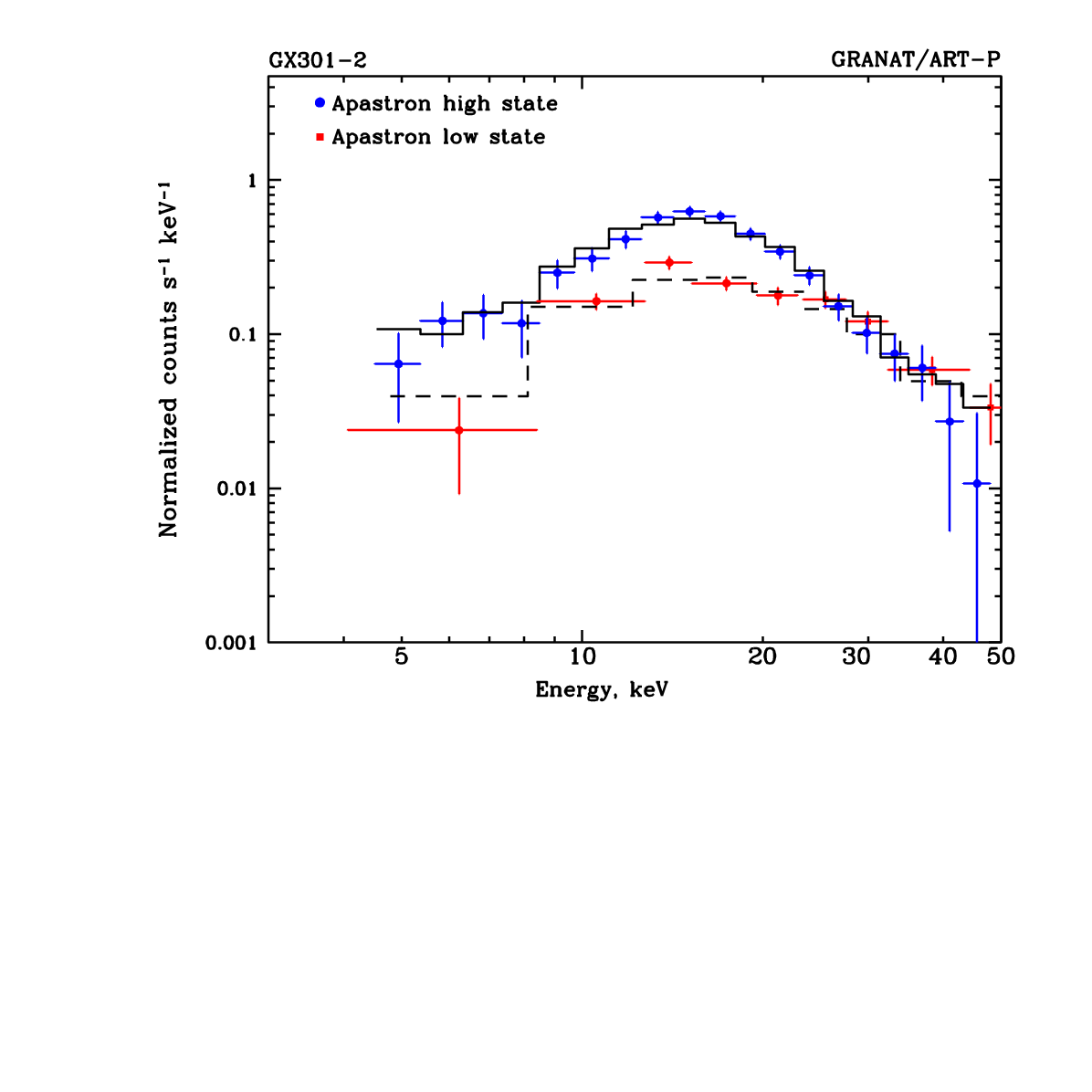}

%\vfill

\caption{
Spectral evolution of the pulsar GX 301-2 near
the apoastron, as inferred from the ART-P data obtained
on February 7, 1992. The normalization of the low-state
spectrum was multipliedby a factor of 2.6. The power-law
fits to the high-state (solid lines) and low-state (dashed
lines) spectra are indicated by histograms. The errors
correspond to one standard deviation.}

\end{figure*}

\pagebreak

\begin{figure*}[t]
\includegraphics[width=14cm,bb=90 140 550 700]{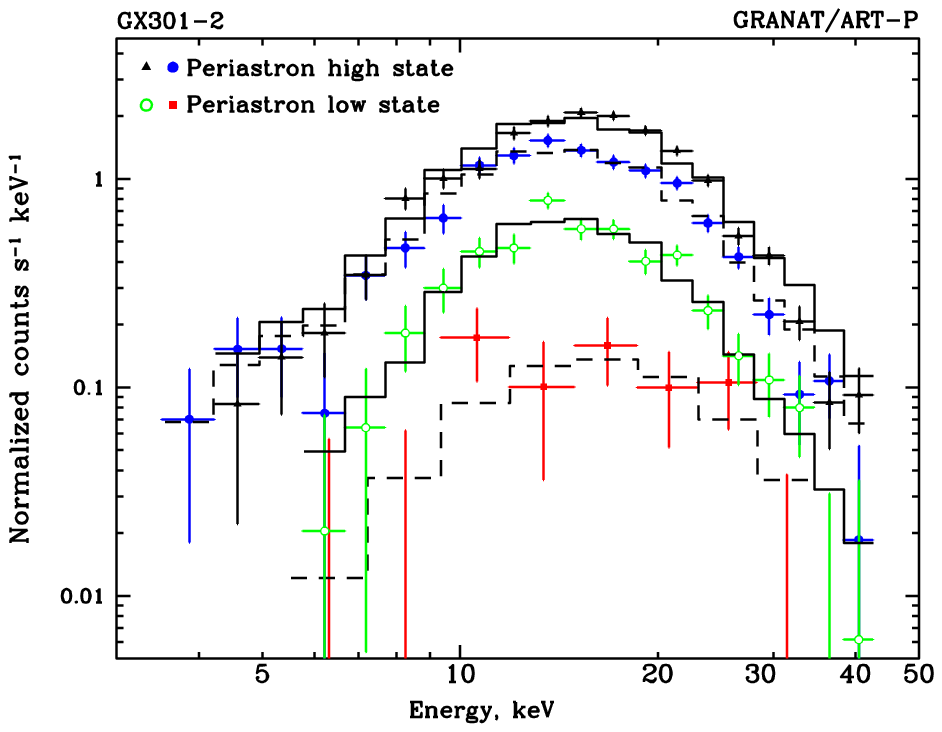}

%\vfill

\caption{
Spectral evolution of the pulsar GX 301-2 near
the periastron, as inferred from the ART-P data obtained
on January 9, 1991. The power-law fits to the high-
state (solid lines) and low-state (dashed lines) spectra
are indicated by histograms. The errors correspondto one
standard deviation.}

\end{figure*}

\newpage

\begin{figure*}[t]
\includegraphics[width=14cm,bb=100 240 500 690,clip]{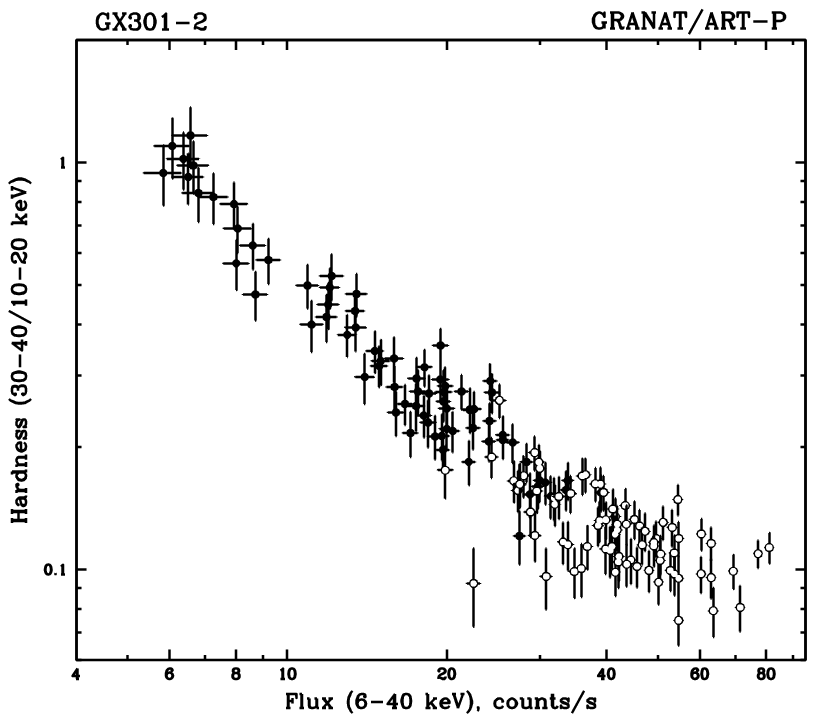}

%\vfill

\caption{Hardness of the source versus its intensity during
periastron passage on February 7, 1992 (filled circles),
and apoastron passage on January 9, 1991 (open circles).}

\end{figure*}

\end{document}